# Weak ferromagnetism and spin glass state with nano-sized nickel carbide

Lin He[1], Yonghua Leng[2], Chinping Chen[1*], Xingguo Li[2*]

[1] Department of Physics, Peking University, Beijing, 100871, People's Republic of China

[2] The State Key Laboratory of Rare Earth Materials Chemistry and Applications, College of Chemistry and Molecular Engineering, Peking University, Beijing 100871, People's Republic of China

Abstract:

$Ni_3C$ nanoparticles of about 40 nm have been studied experimentally to exhibit weak ferromagnetic (FM), spin-glass (SG) and paramagnetic (PM) properties. The freezing temperature of the SG phase at zero applied field is determined as, $T_F^0 \sim 11.0$ K. At $T > T_F^0$, a very weak ferromagnetism has been observed over a PM background. The Curie temperature, $T_C$, is shown to exceed 300 K and the ferromagnetism at 300 K is determined as about 0.02 emu/g ($\sim 6.7 \times 10^{-4}$ $\mu_B$ per $Ni_3C$ formula unit) by subtracting the background paramagnetism. An anomalous dip appears in the temperature dependent coercivity, $H_C(T)$, near the freezing temperature, $T_F^0$. It reflects a distortedly reduced coercivity in the $M(H)$ hysteresis loop measured at $T \sim T_F^0$ with the applied sweeping field around $H = 0$. This is attributable to the exchange coupling effect between the SG and the weak FM phases. The possible origin of the magnetic moments that account for the observed FM, SG and PM properties is discussed.

## I. Introduction

The synthesizing methods and the properties of nickel carbide ($Ni_3C$) have long been under investigation for several decades [1-4]. Recently, the synthesis and magnetic properties of $Ni_3C$ nanoparticles have become one of the focused points due to the rising interests in nanoscience and nanotechnology. For example, nanoparticles of 40 nm have been synthesized by the thermal decomposition of nickel formate, $Ni(HCOO)_2$ [5], the magnetic properties of 10 nm particles obtained by mechanically alloying method have been studied in details [6]. $Ni_3C$ is believed to be nonmagnetic theoretically due to the strong hybridization of the Ni and C orbitals. However, a weak ferromagnetism estimated of the order of 0.6 emu/g (~0.02 $\mu_B$ per $Ni_3C$ formula unit) at $T = 300$ K has been encountered in experiment [6]. The origin of the observed magnetism has been ascribed to the presence of crystal defects. According to a linear muffin-tin orbital (LMTO) band structure calculation, the C vacancies which generate locally Ni-rich regions are believed to sustain the magnetic moments exhibiting the PM and FM properties. In addition to the PM and FM states, the magnetic SG state has also received much attention with the rising interests in nanoscale magnetic particles. For example, with the 6.5 nm $NiFe_2O_4$ particles [7] and the 10 nm $\gamma$-$Fe_2O_3$ particles [8], a magnetic core-shell model has been proposed to explain the observed magnetic properties. By this model, the magnetic nanoparticles are considered as consisting of a FM core and a surface SG layer.

In this report, we present a detailed experimental study on the magnetic properties of $Ni_3C$ nanoparticles with the size of about 40 nm. The sample exhibits a "surface"

SG behavior at low temperature with $T_F^0 \sim 11.0$ K and a very weak ferromagnetism over a PM background at $T > T_F^0$. The weak ferromagnetism is estimated as 0.02 emu/g ($\sim 6.7 \times 10^{-4} \mu_B$ per $Ni_3C$) at $T = 300$ K. It is smaller than the value, $\sim 0.6$ emu/g, reported previously by more than an order of magnitude [6]. To confirm the presence of the SG phase in the low temperature region, we have performed three different magnetic measurements. First, the freezing temperature $T_F$ has been determined by the field-cooling (FC) and zero-field-cooling (ZFC) dc magnetization measurement. The dependence of $T_F(H_{app})$ on the applied measuring field, $H_{app}$, agrees with the de Almeida–Thouless (AT) line $\Delta T_F = T_F^0 - T_F(H_{app}) \propto H_{app}^{2/3}$ [8-12]. Second, the results of the temperature dependent ac susceptibility measurements support the presence of a SG phase at low temperature. Third, the time dependent measurements of the thermoremanent magnetization (TRM) also reveal the SG property. Interestingly, the temperature dependent coercivity, $H_C(T)$, exhibits an anomalous behavior by showing a dip at the temperature around $T_F^0$. It is attributable to the effect of exchange interaction between the FM and the SG phases. Similar behavior in $H_C(T)$ has been observed with 3 nm Co-Ni-B and Fe-Ni-B nanoparticles reported recently. A Monte Carlo (MC) simulation has ascribed this behavior to the effect of the FM-SG interaction [13].

## II. Sample preparation and characterization

$Ni_3C$ nanoparticles were prepared by the thermal decomposition of $Ni(CHOO)_2$. Detailed synthesis procedure was published in a previous report [5]. In brief, a typical solution of $Ni(CHOO)_2$ with the presence of oleic acid and oleylamine as surfactants was heated to ~529 K in a $N_2$ atmosphere for about 2 hours. The active Ni atoms in the compound were then carbonized by the active carbon contained in the surfactants. Figure 1 shows the XRD pattern for the sample in the log-Y scale. The Bragg peaks are indexed nicely to the crystal planes of nickel carbide [14].

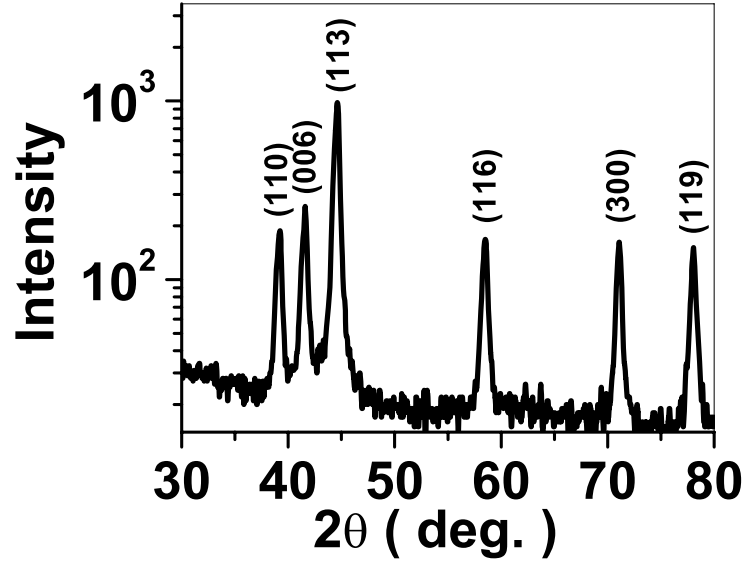

Figure 1: XRD patterns of the sample. The Bragg peaks are conformed to the crystal planes of nickel carbide ($Ni_3C$) [14].

### III. Magnetic measurements and analysis

Magnetic properties of the sample were studied by dc magnetization measurements using a Quantum Design MPMS SQUID magnetometer, and by ac susceptibility measurements using a Quantum Design PPMS system. The sample mass was 4.60 mg in a powder collection.

### a) FC and ZFC dc magnetization measurements

The ZFC curve, $M_{ZFC}(T)$, and FC curve, $M_{FC}(T)$, between 5 K and 300 K are shown in Fig. 2. For the $M_{ZFC}(T)$ measurement, the sample was cooled under a zero applied field from 300 K down to 5 K, and then a field of 90 Oe was applied for data collection in the warming process. For the $M_{FC}(T)$ curve, the procedure was the same as that for the ZFC measurement, except that the sample was cooled in a cooling field of 20 kOe. Two major features are revealed by the ZFC-FC measurements. First, a sharp peak is present on the $M_{ZFC}(T)$ curve at low temperature, $T_F = 10.3$ K, which will be confirmed later as the freezing temperature for a "surface" SG phase. Second, a weak ferromagnetism with $T_C > 300$ K is present over a PM background at $T > T_F$. An amplified view is shown in the inset with a log-Y scale to better reveal the separation bwtween the $M_{ZFC}(T)$ and $M_{FC}(T)$ curves with temperature going up to $T > 300$ K.

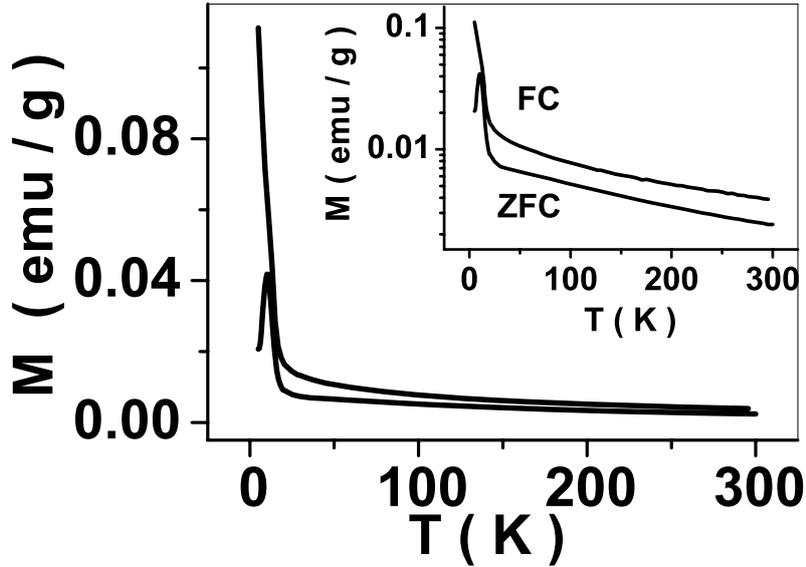

Figure 2: $M_{ZFC}(T)$ and $M_{FC}(T)$ measurements recorded in $H_{app} = 90$ Oe. The inset shows the same $M_{ZFC}(T)$ and $M_{FC}(T)$ curves in the log-Y scale.

For the confirmation of a surface-like SG phase, the $M_{ZFC}(T)$ curves have been measured by various fields, $H_{app}$, including 90, 200, 500, 800 Oe and 1 kOe. The results are plotted in Fig. 3. The peak position, $T_F(H_{app})$, shifts down progressively with the increasing field, $H_{app}$. The relation between the freezing temperature, $T_F$, and the corresponding external applied field, $H_{app} = H_{AT}$ can be described by the AT equation, see for example Refs 8 to 12,

$$H_{AT}(T_F)/\Delta J \propto \left(1 - \frac{T_F}{T_F^{\,0}}\right)^{3/2}, \qquad (1)$$

where $T_F^{\,0}$ is a fitting parameter representing the freezing temperature in a vanishing magnetic field and $\Delta J$ is the width of the distribution of exchange interaction for the SG phase. The inset shows the figure for $H_{AT}^{2/3}$ versus $T_F$. The error bars, which are barely visible in the inset, represent the uncertainty in determining $T_F$. It is determined by the step size in temperature for the measurement, ~ 0.2 K. The solid line going through the points is for the fitting result by the AT equation. According to the analysis, the $H^{2/3}$ dependence reflects a "surface" SG behavior, which is different from the $H^{1/2}$ dependence for a "volume" SG phase as is discussed in [8] and [12]. The zero field freezing temperature is then obtained as $T_F^{\,0} = 11.0$ K by the extrapolation of the AT line to the $T_F$ axis ($H_{AT} = 0$), as shown in the inset of Fig. 3.

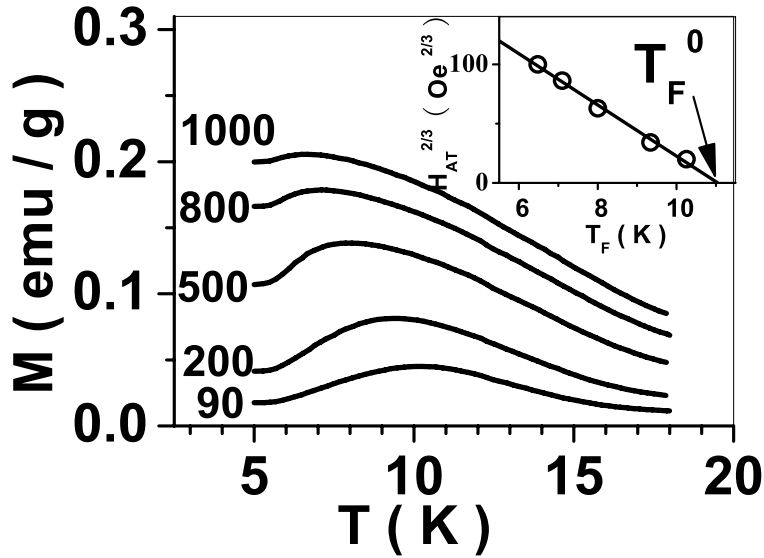

Figure 3: $M_{ZFC}(T)$ data recorded in $H_{app}$ = 90, 200, 500, 800 Oe and 1 kOe. The inset shows the field versus the freezing temperature. The solid line represents the fitting result by the de Almeida–Thouless equation.

**b) ac susceptibility measurements**

A further investigation on the frequency response of the low temperature SG phase has been performed by the ac susceptibility measurements for $\chi_f(T) = \chi'_f(T) + i\chi''_f(T)$. Figure 4a shows the real part of the ac susceptibility, $\chi'_f(T)$, measured by three different frequencies, $f$ = 97, 987, and 9987 Hz with the ZFC mode. For this measurement, the sample was first cooled under a zero applied field from 300 K down to 5 K. Then, a probing ac magnetic field, with the amplitude $\Delta H(f)$ = 10 Oe and the frequency, $f$, was applied for the data collection in the warming process. The dc biased field was set to zero, $H_{dc}$ = 0, during the measurement. The freezing temperature, $T_F(f)$, is identified as the maximum in the $\chi'_f(T)$ curve. It increases with the increasing frequency, $f$. It is noted that a superparamagnetic (SPM) phase would cause a peak in the $M_{ZFC}(T)$ curve by a dc magnetization measurement and, by an ac susceptibility measurement, the peak position for the blocking temperature, $T_B$, also shifts with the measuring frequency, $f$. However, the frequency response of the peak in the $\chi'_f(T)$ curve is different for the SPM phase from the SG phase. The criterion to discern these two magnetic phases can be described by the empirical equation, $\phi = \Delta T_F/[T_F(f)\log(f)]$, in which $\Delta T_F(f) = T_F(f) - T_F^0$ is the shift in the freezing temperature for the SG phase (or the blocking temperature for the SPM phase) from the value of $T_F^0$ = 11.0 K. For a SG phase, the $\phi$ value falls within the range of 0.004 to 0.018, whereas for a SPM phase, a much larger value of $\phi \sim 0.3$ is observed [15,16]. The $\phi$ value with the present Ni$_3$C sample is calculated as 0.012. It falls within the proper range for a SG phase. Figure 4b shows the imaginary part of the ac susceptibility, $\chi''_f(T)$, expressed in the same unit of $10^{-4}$ emu/g.Oe as for $\chi'_f(T)$. Its magnitude is about a tenth of the real part $\chi'_f(T)$. The solid curves in the figure are the fitting results by a polynomial to go through the experimental data points. As the measuring frequency increases, the scattering level of the data points from the fitting curves becomes smaller and the peak position goes up higher in temperature. This is expected for the dynamical behavior of a SG phase.

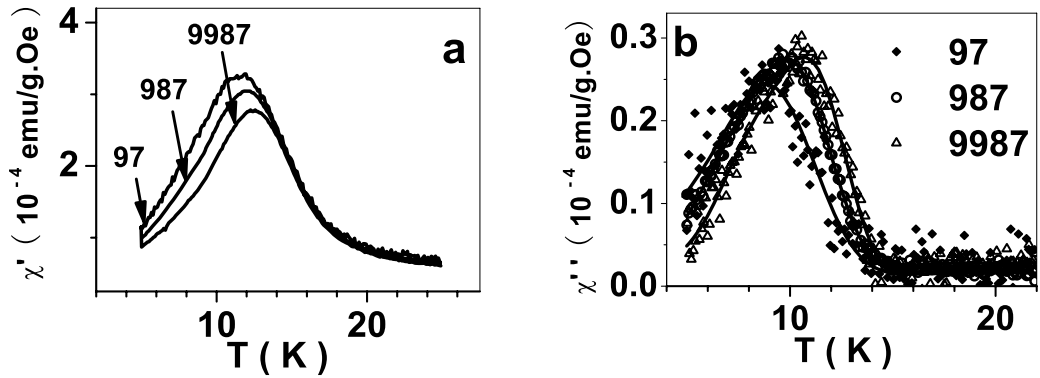

Figure 4: ac susceptibility measurements by ZFC mode using three different frequencies, 97, 987, and 9987 Hz. (a) Real part of the ac susceptibility. (b) Imaginary part of the ac susceptibility. The solid curves are fitting results by polynomial function.

For a SG phase, two different descriptions have been proposed theoretically for its dynamical behavior near the freezing temperature. For the first one, it assumes that a phase transition occurs at the freezing temperature, $T_F^0$. Near the transition point, $T_F^0$, a critical behavior is present for the temperature dependence of the relaxation time, $\tau$, described by the following equation [17-19],

$$\tau = \tau_0 \left(\frac{T_F}{T_F^0} - 1\right)^{-zv}. \qquad (2)$$

In the expression, $1/\tau = f$ is the frequency of the ac measurement, $\tau_0$ is a constant typically in the range of $10^{-9}$-$10^{-13}$ sec and $zv$ is the critical exponent. By fitting the peak position, $T_F(f)$, using Eq. (2), the parameters are obtained as $\tau_0 = 1\times10^{-10}$ sec, $T_F^0$ = 11.0 K and $z\nu$ = 6.5. The value of $T_F^0$ is the same as that obtained by the AT analysis on the dc magnetization measurements, and $\tau_0$ falls in the proper range for a SG phase. The critical exponent obtained here is close to the value of 5.3 for the $Fe_2O_3$ spin cluster [19]. Generally, the value of the critical exponent ranges from 4 to 10 as pointed out in the previous experiments [18,19]. The second description for the dynamical property of a SG phase considers the freezing transition as a nonequilibrium phenomenon. This is describable by the Vogel-Fulcher law [20], which takes into account the interacting property of the SG clusters,

$$\tau = \tau_0 \exp[\frac{E_a}{k_B(T_F - T_0)}]. \qquad (3)$$

In the above equation, $E_a$ is the energy barrier and $T_0$ is a phenomenological parameter describing the inter-cluster interactions. Obviously, Eq. (3) reduces to the Arrhenius law, $\tau = \tau_0 \exp(\frac{E_a}{k_B T_F})$, if the interaction between the SG clusters is negligible. In Fig. 5, $T_F(f)$ versus $\ln(1/f)$ is plotted by the solid circles along with the fitting curve by Eq. (3). From the fitting, we obtained $E_a/k_B$ = 38.3 K, $T_0$ = 9.6 K and $\tau_0 = 1\times10^{-10}$ sec. The value of $T_0$ = 9.6 K > 0 gives an evidence for the presence of interacting SG clusters within the sample. Similar result has also been observed with the amorphous $Fe_2O_3$ nanoparticles of 12 nm in size [19]. In this case, the parameter of interaction obtained by fitting the experimental data using the Vogel-Fulcher law is as large as $T_0$ ~ 34 K. For $\tau_0$, it is obtained from the present experiment as $1\times10^{-10}$ sec, which is equal to the value determined by the first description discussed above. If the experimental result with our sample is interpreted by the Arrhenius law, *i.e.* $T_0 = 0$, it will result in an unphysical outcome with $\tau_0 = 3.7 \times 10^{-32}$ sec. This is also encountered by the surface SG phase with the NiO nanoparticle, ~ 5.1 nm, where $\tau_0 = 10^{-39}$ sec is obtained by applying the Arrehenius law for the fitting of the experimental data [21].

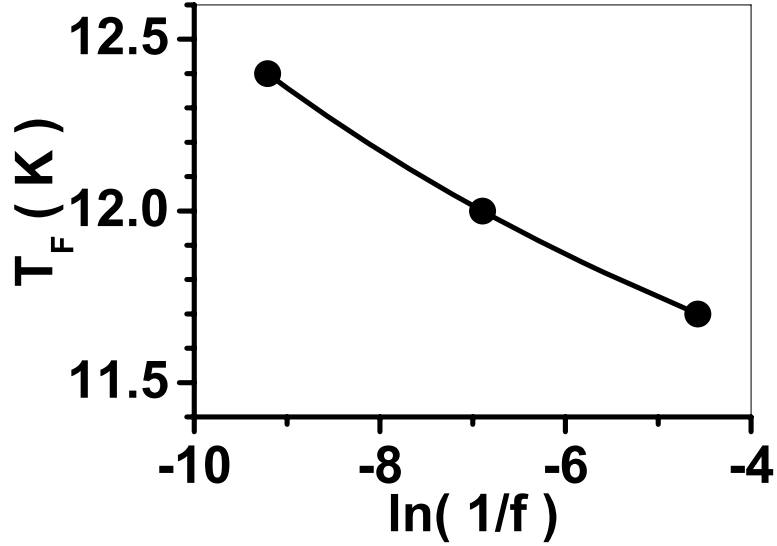

Figure 5: Freezing temperature, $T_F$, versus $\ln(1/f)$, in which $f$ is the measuring frequency in the ac susceptibility measurement and the solid curve is the fitting result by the Vogel-Fulcher law.

| $t_W$ (sec) | $M_{FM}^{TRM}$ (emu/g) | $M_{SG}^{TRM}$ (emu/g) | $\tau$ (sec) | n |
|---|---|---|---|---|
| 10 | 0.0702 | 0.0217 | 4296 | 0.55 |
| 100 | 0.0746 | 0.0183 | 3268 | 0.533 |
| 1000 | 0.0755 | 0.0183 | 3147 | 0.527 |

Table 1: Parameters determined by the fit using Eq. (4) for the thermoremanent magnetization at $T = 5$ K with different waiting time $t_W$.

**c) Thermoremanent magnetization measurements**

The SG behavior has been further studied by the TRM measurements, $M_{TRM}(t)$, in a time scale of $1.1\times10^4$ sec. For the measurements, the sample was first cooled in an applied field of 10 kOe from 50 K > $T_F^0$, down to 5 K < $T_F^0$. With a certain period of waiting time $t_W$ after the temperature was stabilized at 5 K, the field was reduced to zero and the magnetization was recorded at 5 K in zero field as a function of time. Figure 6 shows $M_{TRM}(t)$ measured at 5 K with $t_W$ = 10, 100, and 1000 sec. The aging effect, *i.e.* the dependence of $M_{TRM}(t)$ upon the waiting time $t_W$, has demonstrated that the sample was in a metastable state [16,22]. The time dependent magnetization corresponding to different waiting time, $t_W$, can be described by the stretched exponential function,

$$M_{TRM}(t) = M_{FM}^{TRM} + M_{SG}^{TRM}\exp[-(\frac{t}{\tau})^{1-n}], \quad (4)$$

where $M_{FM}^{TRM}$ is a time independent parameter for the FM phase, and $M_{SG}^{TRM}$ is for the SG component which is responsible for the observed relaxation behavior [13,16,19,22]. For a SG system, the parameter $n$ depends only on the measuring temperature $T$, while the time constant $\tau$ and $M_{SG}^{TRM}$ show a large variation with the waiting time $t_W$ [16,19]. The fitting results by Eq. (4) with $t_W$ = 10, 100, and 1000 sec are shown in Fig. 6 by the solid curves. The parameters obtained from the fitting are listed in Tab. 1. The values of $M_{FM}^{TRM}$ and $n$ change only about 5%, while $M_{SG}^{TRM}$ varies about 19% and the time constant $\tau$, about 36% with $t_W$ changing from 10 to 1000 sec. According to this analysis, $M_{FM}^{TRM}$ is obtained as about 0.075 emu/g at $T$ = 5 K. Although determined at a much lower temperature, this value is on the same order of magnitude as that determined by the *M(H)* loop at *T* = 300 K, which is about 0.02 emu/g presented in the next subsection. On the other hand, $M_{SG}^{TRM}$ determined by this measurement is about 0.02 emu/g, which is smaller by an order of magnitude than the value determined from the *M(H)* loop at *T* = 5 K, ~ 0.4 emu/g discussed also in the next subsection. This is not surprising since $M_{SG}^{TRM}$ depends heavily on the magnitude of the cooling field and the waiting time $t_W$.

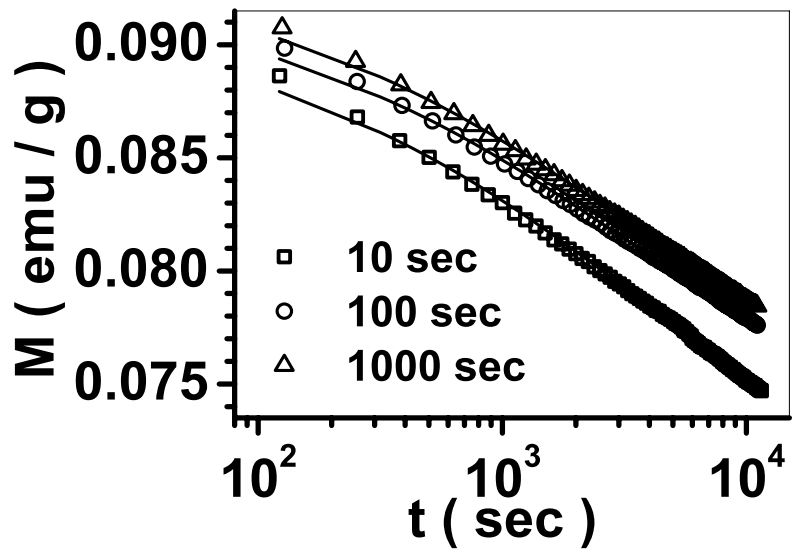

Figure 6: Thermoremanent magnetization, $M_{TRM}(t)$, at *T* = 5 K with the waiting time of $t_W$ = 10, 100, and 1000 sec. The solid curves are the fitting results by Eq. (4).

**d) Field dependent magnetization measurements**

The *M(H)* measurements have been performed at *T* = 5, 7, 8.5, 10, 12, 13.5 15, 20, 50, 100, 150, 200, 250, and 300 K. Four representative hysteresis loops measured at *T* = 5, 12, 20 and 300 K are plotted in Fig. 7. These curves do not saturate even at the maximum applied field of 30 kOe, as shown in the insets of Figs. 7a, 7b, 7c, and 7d. This is a typical behavior for a magnetic system with a PM or a SG phase. In Fig. 7a, the open loop measured at $T = 5$ K $< T_F^0$ is much more pronounced in comparison with the ones measured at $T > T_F^0$, e.g. at 20 K shown in Fig. 7c and 300 K shown in Fig. 7d. This indicates an important contribution of the SG phase to the irreversible behavior of the $M(H)$ curve in response to the sweeping applied field. Similar enhancement in the hysteresis loop attributed to the surface SG phase at $T < T_F^0$ has been reported recently with $(Cu)_{core}/(Cu_2O+CuCl+minuteCuO)_{shell}$ nano-composite [23] and Ni nanochains [12]. On the other hand, Fig. 7b shows a much reduced coercivity in the low field region at $T = 12$ K. The coercivity almost vanishes, ~ 8 Oe, in comparison with the values at $T = 5$ K, *i.e.* $H_C(5$ K$) \sim 295$ Oe, and $T = 20$ K, *i.e.* $H_C(20$ K$) \sim 60$ Oe. It is even smaller than the value measured at 300 K, *i.e.* $H_C(300K) = 32$ Oe, as shown in Fig. 7d. It actually exhibits a distorted shape of the $M(H)$ curve around $H = 0$, resulting in the vanishing coercivity. This anomalous behavior has been reproduced on the same sample after the sample storage at room temperature for 5 months. The solid curve in Fig. 7b is for the first measurement, while the open circles are for the second one. It is known that the coercivity of magnetic nanoparticles would reveal a feature of reduced magnitude arising from the interparticle dipolar coupling [24] or from the magnetization vortex formation within the nanoparticle [25]. However, this is not the case with the present sample. Similar "distorted" loops around $H = 0$ with an appreciable reduction of coercivity have also been observed near the freezing temperature with the 3 nm Co-Ni-B and Fe-Ni-B nanoparticles [13,26], see Fig. 2b in Ref. 13. This anomalous behavior has been ascribed to the SG-FM exchange interaction according to a simple MC simulation. We believe that this is the mechanism responsible for the severe reduction of coercivity in the $M(H)$ curve measured at $T = 12$ K which is slightly above the freezing temperature, $T_F^0 = 11.0$ K, of the SG phase.

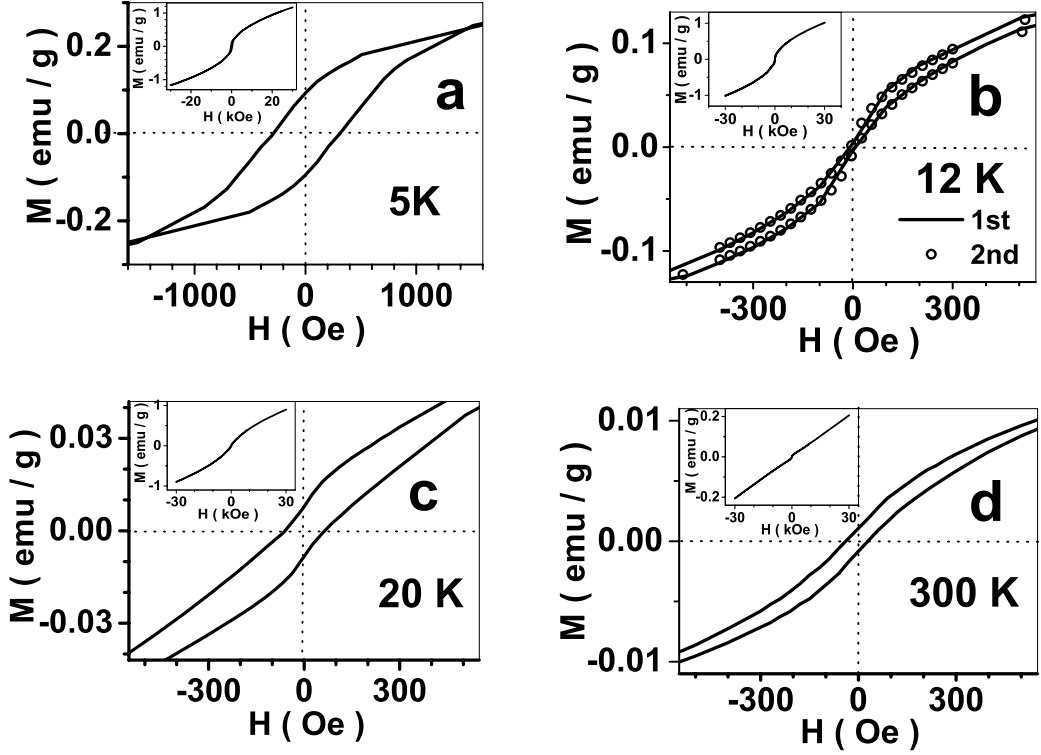

Figure 7: Hysteresis loops measured at $T = 5, 12, 20$ and $300$ K in the low field region. The insets show the $M(H)$ curves in the entire range of the applied field, $-30$ kOe $< H <$ 30 kOe.

Figure 8 shows the temperature dependence of coercivity, $H_C(T)$, and saturation magnetization, $M_S(T)$, determined from the experimentally measured $M(H)$ curves. The saturation magnetization, $M_S$, for the FM phase, which even includes the contribution of the SG phase at $T < T_F^0$, is determined by the extrapolation of the high field linear part of the $M(H)$ curve to the axis of $H = 0$. It increases from about 0.02 emu/g at $T = 300$ K to roughly 0.5 emu/g at $T = 5$ K. The dramatic increase by more than one order of magnitude in the saturation magnetization at low temperature indicates that the magnitude of magnetic moment is much more significant corresponding to the SG phase than the FM phase. The magnetization of the SG phase at $T = 5$ K is then estimated as about 0.4 emu/g by subtracting the FM component determined by the TRM measurement, $M_{FM}^{TRM}$ ~0.075 emu/g. The anomalous dip in $H_C(T)$, showing up near the freezing temperature as shown in Fig. 8, is a reflection of the distortedly reduced coercivity around $H = 0$ in the $M(H)$ curve. According to the MC simulation, the exchange interaction between the FM core and the SG surface layer of a nanoparticle with a magnetic core-shell structure is essential to result in such an anomalous behavior in $H_C(T)$ [13,26]. In the calculation, the spins in the surface SG layer is randomly oriented and nearly defreezing at $T \sim T_F^0$. These spin moments would then contribute to demagnetize the core ferromagnetism via the SG-FM exchange coupling as $H$ approaches zero. Hence, a severely reduced coercivity is observed around $H = 0$. On the other hand, as the temperature goes down further to $T < T_F^0$, the spin moments of the SG phase are highly frozen. The SG-FM coupling actually exerts a strong pinning effect on the FM phase to cause an increase in the coercivity.

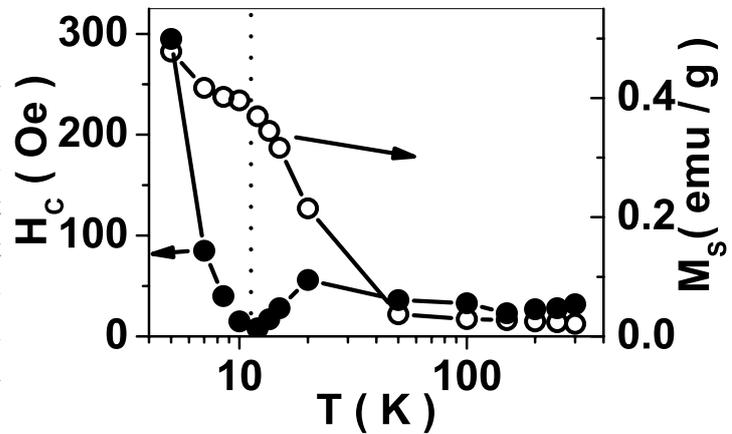

Figure 8: Temperature dependence of coercivity, $H_C(T)$, obtained from the hysteresis loops measured at different temperatures. The right Y axis shows the temperature dependence of the saturation magnetization, $M_S(T)$, obtained by the method of extrapolation.

## IV. Discussion

In the present work, interesting and complicated magnetic properties have been observed with a powder collection of 40 nm $Ni_3C$ nanoparticles. The sample reveals a significant "surface" SG behavior with $T_F^0$ = 11.0 K and a very weak FM phase, about 0.02 emu/g at $T$ = 300 K with $T_C$ > 300 K. According to the analysis on the $M(H)$ measurement at $T$ = 5 K, the magnetization attributed to the SG phase is estimated as about 0.4 emu/g ($1.3\times10^{-2}$ $\mu_B$ per $Ni_3C$ formula unit). It is stronger than the ferromagnetism at the same temperature, ~ 0.075 emu/g, determined from the TRM measurement. In addition, a significant PM background has been revealed without any evidence for the presence of a SPM phase. The SG phase is shown to be interacting according to the Vogel-Fulcher analysis by Eq. (3). More interestingly, the weak FM and the significant SG phases show a sign of exchange interaction between them, in similarity to a previously published result for the 3 nm Co-Ni-B and Fe-Ni-B nanoparticles [13,26]. The possible origin of the magnetic moments which are responsible for the observed magnetic properties with the present $Ni_3C$ nanoparticles is not a trivial issue to resolve and will be discussed in the next few paragraphs.

There are two possibilities for the origin of the magnetic moments causing the observed weak ferromagnetism, the SG and the PM states. The first is by the Stoner type itinerant magnetism arising from the presence of C vacancies. It is shown by the LMTO band structure calculation that the origin of the magnetic moments with the $Ni_3C$ is attributable to the presence of locally Ni-rich regions caused by the C vacancies. Weak magnetic moments appear due to the shift of density of state (DOS) near the Fermi surface for Ni atoms locating close to the defects [6]. A Stoner-typed itinerant magnetic moment rather than a localized moment is then induced. Two major features of the itinerant moments, which are usually distinctive from the localized spin moments, are the spatially extended nature and the smallness in magnitude. An alternative scenario to account for the presence of the magnetic moments is by the picture of localized magnetic moments attributed to the unreacted Ni atoms or Ni nanoclusters left over from an incomplete carbonization process. Each unreacted Ni atom contributes a localized magnetic moment of about 0.606 $\mu_B$ estimated from the bulk value.

Despite the negative result in the XRD analysis, the presence of unreacted Ni atoms or nanoparticles can not be completely ruled out. However, a number of difficulties exist by the picture of localized magnetic moments to explain the weak ferromagnetism. A concentration of Ni impurity, ~ $3.6 \times 10^{-4}$ g-Ni/g-Ni$_3$C, with a moment of 0.606 $\mu_B$ per Ni atom is enough to explain the observed FM properties at room temperature, ~ 0.02 emu/g. By the assumption that there is an incomplete carbonization region within each Ni$_3$C nanoparticle, then the average size of the residual Ni nanocluster embedded in each Ni$_3$C nanoparticle is estimated as 2.8 nm. This is much smaller than the coherence length of Ni, ~ 25 nm [27,28], and is qualified as a Stoner-Wohlfarth (SW) nanoparticle. A SPM property with a blocking temperature, $T_B$, is therefore expected. The volume of the unreacted Ni nanocluster is calculated as, $V_{Ni} = (4\pi/3)(1.4\text{nm})^3$ ~ 11.5 nm$^3$. By the assumption that the demagnetization factor of the Ni nanocluster is $\Delta N = 0.1$, the shape anisotropy is calculated as $K_{\text{shape}} = (1/2)M_S^2 \Delta N = 1.4 \times 10^4$ J/m$^3$, by using the bulk $M_S$ value of Ni. The blocking temperature is then estimated by the expression, $T_B = K_{\text{shape}}V_{\text{Ni}}/25k_B$, as 0.48 K. For a particle of spherical shape, the demagnetization factor is usually much smaller than 0.1. The blocking temperature is expected even lower. In addition, the Curie temperature of a 2.8 nm Ni particle would reduce from the bulk value of 631 K down to 130 K according to the finite size effect [29]. Hence, $T_B$ is expected further reduced from 0.48 K accordingly. At $T > T_B$, the Ni nanocluster is expected to show SPM behavior [27]. The experimentally observed weak ferromagnetism at 300 K by the FC-ZFC and $M(H)$ measurements is therefore inconsistent with this picture. On the other hand, for a Ni nanoparticle or a nanocluster embedded inside the Ni$_3$C particle to exhibit ferromagnetism at $T > 300$ K, the diameter is calculated to be at least 21 nm by assuming $\Delta N = 0.1$ with the finite size effect [29] and the temperature dependent shape anisotropy effect of Ni accounted for [27]. If the possibly incomplete reaction process leaves the unreacted Ni in the form of a large particle with the diameter $D > 21$ nm, then, it is difficult to interpret the appearance of the dip showing up in $H_C(T)$ around the freezing temperature by a simple magnetic core-shell structure of the Ni nanoparticles. For example, chains of Ni nanoparticles, about 50 nm in diameter, have been studied in a previous report [12]. The magnetic core-shell model has reasonably explained the observed magnetic behavior. However, not a dip has been observed in $H_C(T)$. In order for the FM-SG coupling effect to be manifested, the FM component should be relatively weak in comparison with the SG phase so that the FM signal does not obscure the FM-SG coupling effect [13]. According to the above discussion, the origin of the detected weak ferromagnetism is unlikely coming from the localized moments of unreacted Ni atoms. Rather, it is attributable to the carbon-vacancy-induced residual moments.

For the significant SG phase, it is difficult to apply the mechanism of itinerant magnetism to explain its presence in the sample owing to the randomly oriented nature of the magnetic moments trapped by a random magnetic potential. Instead, it is more likely attributed to the localized moments of unreacted Ni atoms randomly distributed within the $Ni_3C$ nanoparticles. This is possible with a trace amount of unreacted Ni embedded in the nonmagnetic $Ni_3C$ matrix. The magnetization of 0.4 emu/g estimated at $T = 5$ K for the SG phase can be accounted for by the magnetic moment of about $1.3\times10^{-2}$ $\mu_B$ per $Ni_3C$ formula unit. The corresponding Ni impurity is on the level of about $7.2\times10^{-3}$ g-Ni/g-$Ni_3C$. From the nature of the SG phase, which is more surface-like rather than volume-like according to the analysis in the present work, a possible scenario is proposed to offer a reasonable explanation for the experimentally observed properties of the SG phase and the SG-FM coupling effect. For C vacancies, which are defects of Ni-rich region with very weak itinerant ferromagnetism, the surrounding localized moments of metallic Ni impurities randomly distributing around the defects would exhibit SG properties. This would naturally form regions with "magnetic core-shell structure" within the sample, *i.e.*, a weak ferromagnetic core (the defect of C vacancy) surrounding by a significant SG phase (unreacted metallic Ni atoms). With the extended nature of the itinerant ferromagnetism, the SG and the FM phases have large interfacial or contact area for the significant exchange coupling effect to occur. Owing to the weakness of the ferromagnetism, the SG-FM coupling effect is not obscured and is revealed by the $M(H)$ measurements near the freezing point of the SG phase.

For the PM background which shows up in the FC-ZFC $M(T)$ curves and in the high field region of the $M(H)$ curves, the corresponding magnetic moments is more likely to originate from the localized moments of Ni atoms for two reasons. First, the moment arising from the C vacancy is small in nature. Second, the localized moments of the SG phase will exhibit a behavior of irreversible field response below and a PM property above the freezing temperature. The nonsaturating feature of the magnetization in the high field region of a $M(H)$ curve for a magnetic nanoparticle has been ascribed by a numerical MC simulation to the disordered surface spin moments with a magnetic core-shell structure [30]. This property has been observed in many experiments as well, for example, with the 6.5 nm $NiFe_2O_4$ particles [7] and the 10 nm $\gamma$-$Fe_2O_3$ particles [8].

Although the magnetic core-shell model proposed to explain the magnetic properties in the present experiment is plausible, more experimental evidences are still in need for a further confirmation. In particular, direct experimental evidence on the structure characterization for the defect state of C vacancies is highly interesting in the future. Also, the properties of the SG-FM exchange interaction are an interesting subject worthy of further study in addition to the AFM-FM exchange coupling effect which has already received more attention [31].

**V. Conclusion**

We have carried out an experimental investigation on the magnetic properties of Ni$_3$C nanoparticles, about 40 nm in diameter. The $M(T)$, $M(H)$, $\chi_f(T)$, and $M_{TRM}(t)$ measurements confirm the presence of a very weak ferromagnetism, about 0.02 emu/g at $T = 300$ K with the Curie temperature $T_C > 300$ K, and an interacting SG phase with the freezing temperature at $T_F^0 = 11.0$ K. The magnetic moment detected for the weak ferromagnetism is attributable to the presence of C vacancies having the Stoner type itinerant moments, and for the SG phase, it is ascribable to the random distribution of localized moments possibly arising from the unreacted Ni atoms. A model of "magnetic core-shell structure" is proposed to explain the observed magnetic properties, with the SG phase surrounding the weak FM cores formed of the C vacancies within the Ni$_3$C nanoparticles. The anomalous dip showing up in $H_C(T)$ at the temperature around $T_F^0$ is then ascribed to the SG-FM exchange coupling effect. The present studies are interesting for the understanding on the magnetic properties not only of the Ni$_3$C nanoparticle, but also of any material that is theoretically nonmagnetic with ideal crystal structure. Further studies remain interesting to show a more detailed picture on the origin of the magnetism with Ni$_3$C and to explore the properties of SG-FM exchange coupling for any material having both of the FM and magnetic SG phases.

**Acknowledgement**

The authors acknowledge the National Natural Science Foundation of China (Grant Nos.10335040, 20221101, and 20671004).